\documentclass[english]{article}
\usepackage[T1]{fontenc}
\usepackage[latin9]{inputenc}
\usepackage{amstext}
\usepackage{cancel}
\usepackage{graphicx}
\usepackage{esint}
\usepackage{babel}
\begin{document}
\title{\textbf{\large{}Nucleon Resonance Masses from QCD Sum Rules}}
\author{\textbf{\normalsize{}Nasrallah F. Nasrallah$^{(a)}$, Karl Schilcher}\textbf{$^{(b,c)}$}\\
{\normalsize{}$^{(a)}$ Faculty of Science, Lebanese University, Tripoli
1300, Lebanon }\\
{\normalsize{}$^{(b)}$ Institut fur Physik, Johannes Gutenberg-Universitat}\\
{\normalsize{}Staudingerweg 7, D-55099 Mainz, Germany}\\
{\normalsize{}$^{(c)}$Centre for Theoretical Physics and Astrophysics}\\
{\normalsize{}University of Cape Town, Rondebosch 7700, South Africa}\\
\thanks{Supported in part by NRF (South Africa) and Alexander von Humbold
Foundation (Germany)}}
\maketitle
\begin{abstract}
\noindent A method used previously to calculate the masses of the
vector mesons is extended to the calculation of the nucleon resonances.
The method is based on the choice of integration kernels which eliminate
the unknown parts of the spectrum.

We obtain remarkably stable results in a wide range of $R$, the radius
of the integration contour in the complex plane. Agreement with experiment
is good.
\end{abstract}
\newpage{}

\section{QCD sum rules and nucleon resonance masses}

QCD sum rules based on the Borel transformation of the operator product
expansion (OPE) were applied to the calculation of low energy properties
of hardons starting with the pioneering paper of Shifman, Vainshtein,
and Zakharov \cite{SVZ}. The method was soon afterwars applied to
the nucleon by B. L. Ioffe \cite{Ioffe2},\cite{Ioffe}. An alternative
more general approach is based on finite energy sum rules (FESR) \cite{FESR},\cite{Dom}.
Most recently was applied we applied our approach to the calculation
of all vector meson masses \cite{Nasrallah}. We here extend our work
to the calculation of the masses of the nucleon resonances.

Consider the nucleon correlator
\[
\Pi(t=q^{2})=i\int dxe^{iqx}\left\langle 0\left\vert T\eta(x)\eta(0)\right\vert 0\right\rangle 
\]
where
\[
\eta=\varepsilon^{abc}(u_{a}C\gamma u_{b})\gamma_{5}\gamma_{\lambda}d_{c}
\]
is the nucleon current proposed in \cite{Ioffe2}. In a previous work
\cite{Nasrallah1} we studied this correlator to calculate the mass
of the nucleon. We used FESR with polynomial kernels designed to eliminate
the contribution of the nucleon continuum. This continuum consists
mainly of the resonances $N^{+}(1440)$, $N^{-}(1535)$, $N^{-}(1650)$,
$N^{+}(1710)$, $N^{+}(1880)$, $N^{-}(1895)$, $N^{+}(2100)$. Here
we shall use a slightly modified integration kernel to calculate the
masses of nucleon recurrences beginning with the Roper resonance $N^{+}(1440)$.

The amplitude $\Pi(t)$ can be decomposed as
\begin{equation}
\Pi(t)=\cancel{q}\Pi_{1}(t)+\Pi_{2}(t)\label{3.2}
\end{equation}
We shall work with $\Pi_{2}(t)$. At low energies
\begin{equation}
\Pi_{2}(t)=-\lambda^{2}m_{N}\frac{1}{(t-m_{N}^{2})}+...\label{3.3}
\end{equation}
where $\lambda$ is the coupling of the current to the nucleon.

At high energies
\begin{equation}
(2\pi)^{4}\Pi_{2}(t)=B_{3}t\ln(-t)+\frac{B_{7}}{t}+\frac{B_{9}}{t^{2}}\label{3.4}
\end{equation}
with \cite{Nasrallah1} 
\begin{eqnarray}
B_{3} & = & 4\pi^{2}\left\langle \bar{q}q\right\rangle (1+\frac{3}{2}a_{s})=-0.908\text{ GeV}\label{3.5a}\\
B_{7} & = & -\frac{4\pi^{2}}{3}\left\langle \bar{q}q\right\rangle \left\langle a_{s}G^{2}\right\rangle =0.034\text{ GeV}^{7}\label{3.5b}\\
B_{9} & = & -(2\pi)^{6}\frac{136}{81}\left\langle \bar{q}q\right\rangle ^{3}a_{s}=0.083\text{ GeV}^{9}\label{3.5c}
\end{eqnarray}
where we use for $a_{s}=\frac{\alpha_{s}}{\pi}=0.10$, the quark condensate
$\left\langle \bar{q}q\right\rangle =-0.02$ GeV$^{3}$ and the gluon
condensate $\left\langle a_{s}G^{2}\right\rangle =0.013$ GeV$^{4}$.
\begin{figure}[h]
\begin{centering}
\includegraphics[width=0.7\textwidth]{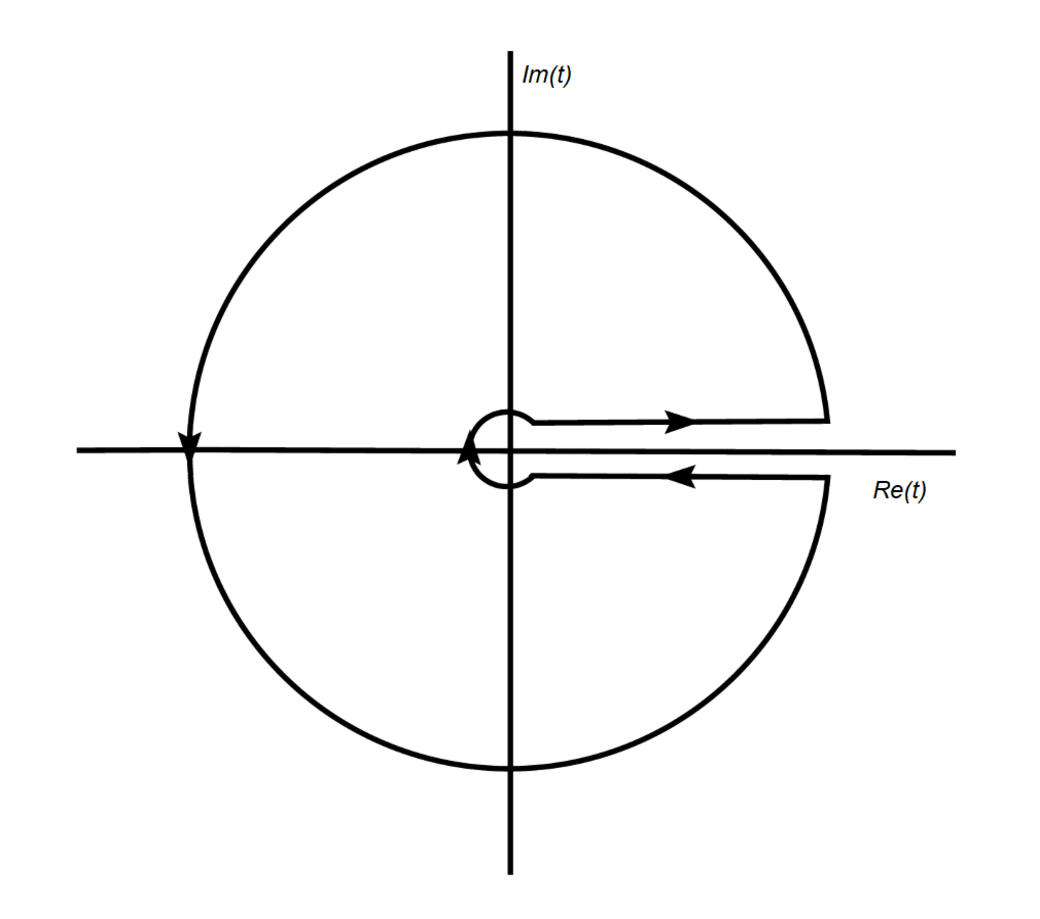}
\par\end{centering}
\centering{}Fig. 1: Integration contour of FESR
\end{figure}

The basis of FESR is Cauchy's theorem applied to the contour $C$
of Fig. 1 which implies
\begin{equation}
-\lambda^{2}m_{N}P(m_{N}^{2})=\frac{1}{\pi}\int_{cut}^{R}dt\,P(t)\textrm{Im}\Pi^{\text{exp}}(t)+\frac{1}{2\pi i}\oint_{\left\vert t\right\vert =R}dt\,P(t)\,\Pi^{\text{QCD}}(t)\label{27}
\end{equation}
where $P(t)$ is an entire function, e. g. a polynomial. Over the
circle of large radius $R$ the correlator $\Pi(t)$ has been replaced
by its QCD expression. The principal unknown in Eq.(\ref{28}) is
the integral over the cut, i. e. over the higher nucleon resonances
with mass $m_{N_{i}}^{2}\leq R$. To minimize this integral (before
neglecting it), a judicious choice of the weight-function $P(t)$
has to be made. With the classic choice \cite{SVZ} $P(t)=\exp(-t/M_{0}^{2})$
the Borel variable $M_{0}^{2}$ cannot be chosen too large because
it would minimize the contribution of the\ nucleon. Also $M_{0}^{2}$
cannot be too small because the unknown condensates in Eq.(\ref{3.4})
would explode. It was hoped in \cite{SVZ} that a region of stability
at an intermediate $M_{0}^{2}$ can be found. This can be shown to
be not the case \cite{Ioffe}. In our FESR approach \cite{Nasrallah1}
we take $P(t)$ to be a polynomial
\begin{equation}
P(t)=\sum_{n=0}^{n_{\text{max}}}c_{n}t^{n}\label{28}
\end{equation}
It is clear that the order $n_{\text{max}}$ cannot be chosen arbitrarily
high because of the contribution of unknown higher condensates. To
test the method we start with the nucleon proper. We choose

\begin{equation}
P(t,R)=(1-\frac{t}{m_{N_{1}}^{2}})(1-\frac{t}{R})=1-a_{1}t-a_{2}t^{2}
\end{equation}
where 
\[
a_{1}(R)=\frac{1}{m_{N_{1}}^{2}}+\frac{1}{R},\ a_{2}(R)=-\frac{1}{m_{N_{1}}^{2}R}
\]
Eq.(\ref{27}) then becomes
\begin{eqnarray}
(2\pi)^{4}\lambda^{2}m_{N}^{3}P(m_{N}^{2}) & = & -B_{3}I_{2}(R)-B_{9}+\Delta_{2}\label{1a}\\
(2\pi)^{4}\lambda^{2}m_{N}^{2}P(m_{N}^{2}) & = & -B_{3}I_{1}(R)-B_{7}+a_{1}B_{9}+\Delta_{1}\label{2a}
\end{eqnarray}
where 
\begin{equation}
I_{1}=\int_{0}^{R}dt\,t^{2}P(t,R),\ \ \ \ \ I_{2}=\int_{0}^{R}dt\,t\,P(t,R)\text{ }\label{3.7}
\end{equation}
and
\begin{equation}
\Delta_{1}=-(2\pi)^{4}\int_{\textrm{thr}}^{R}dt\,P(t)\textrm{Im}\Pi_{2}(t),\ \ \ \ \Delta_{2}=-(2\pi)^{4}\int_{\textrm{thr}}^{R}dt\,t\,P(t)\textrm{Im}\Pi_{2}(t)\label{3.7a}
\end{equation}

$P(t)$ is designed to minimize the contribution of the continuum
in Eq.(\ref{27}), i. e. to minimize the contribution of $\Delta_{1}+\Delta_{2}$
so as to neglect it then. The nucleon mass is obtained from the ratio
(Eq. \ref{1a})/(Eq. \ref{2a}).

There is a wide region of stability $R$ for $I_{1}(R)$ and $I_{2}(R)$
at around $R=3\text{ GeV}^{2}$

Using the sum rule Eq.(\ref{27}) 
\begin{eqnarray}
-(2\pi)^{4}\lambda^{2}m_{N}^{3}P_{1}(m_{N}^{2}) & = & -B_{3}I_{2}(R)-B_{9}-\Delta_{2}\label{1}\\
-(2\pi)^{4}\lambda^{2}m_{N}P_{1}(m_{N}^{2}) & = & -B_{3}I_{1}(R)-B_{7}+a_{1}B_{9}-\Delta_{1}\label{2}
\end{eqnarray}
where
\begin{equation}
\Delta_{1}(t)=-(2\pi)^{4}\int_{\textrm{thr}}^{R}dt\,tP(t)\textrm{Im}\Pi_{2}(t)\text{ \ \ and \ }\Delta_{2}(t)=-(2\pi)^{4}\int_{\textrm{thr}}^{R}dt\,t^{2}P(t)\textrm{Im}\Pi_{2}(t)\text{ .}\label{3}
\end{equation}

The delta's denote the contribution of the nucleon continuum, they
receive no contribution from the nucleon pole. The aim is to choose
$P(t)$ which minimizes $\Delta_{1}$ and $\Delta_{2}$ so as to neglect
them and obtain $m_{N}$ from Eq.$(1)/(2)$. This was done in \cite{Nasrallah1}
from the choice 
\begin{equation}
P(t)=1-a_{1}t-a_{2}t^{3}=1-0.807t+0.16t^{2}\label{4}
\end{equation}
This choice minimizes the integral $\int_{\textrm{2Ge\ensuremath{V^{2}}}}^{\textrm{3Ge\ensuremath{V^{2}}}}dt\left|P(t)\right|^{2}$

For $R=2.5\text{ GeV}^{2}$ the result was 
\begin{equation}
m_{N}=0.830\pm0.05\text{ GeV}\label{5}
\end{equation}
Note that to obtain the physical mass one has to add the contribution
of the $\Sigma$-term. Using higher moments we get we get information
on the higher resonances

We next proceed with a different choice for $P(t)$ which eliminates
$\Delta_{1,2}$ as well.
\begin{equation}
P^{\prime}(t,R)=(1-\frac{t}{m_{1}^{2}})(1-\frac{t}{R})\label{6}
\end{equation}
which vanishes at the mass of the $m_{1}$, where $m_{1}=m_{N}(1440)$
is the mass of the first exited state (the Roper) where we expect
the bulk of the contribution to the continuum to come from. $R$ will
be determined by stability considerations.

The expression for $m_{N}$ is
\begin{equation}
m_{N}^{2}=\frac{I_{2}^{\prime}(R)+\frac{B_{9}}{B3}}{I_{1}^{\prime}(R)+\frac{B_{7}}{B_{3}}-a_{1}^{\prime}\frac{B_{9}}{B3}}\label{7}
\end{equation}
where
\begin{eqnarray}
I_{2}^{\prime} & = & \int_{0}^{R}dt\,t^{2}P^{\prime}(t,R)\text{ , \ \ }\ \ \ I_{1}^{\prime}=\int_{0}^{R}dt\,t\,P^{\prime}(t,R)\label{8}\\
P^{\prime}(t) & = & 1-a_{1}^{\prime}t-a_{2}^{\prime}t^{2}\text{, \ }\ \ a_{1}^{\prime}=(\frac{1}{m_{1}^{2}}+\frac{1}{R})\ \ \ a_{2}^{\prime}=\frac{1}{m_{1}^{2}R}
\end{eqnarray}
With $m_{N}^{2}$ as an input, Eq.(\ref{7}) and Eq.(\ref{8}) determine
$m_{1}$ as a function of $R$:
\begin{eqnarray*}
R & = & 2.1\text{ GeV}^{2}\ \ \ \ \ m_{1}=1.57\text{ GeV}\\
R & = & 2.4\text{ GeV}^{2}\ \ \ \ \ m_{1}=1.46\text{ GeV}\\
R & = & 3.0\text{ \ensuremath{\text{GeV}^{2}\ \ \ \ \ }}m_{1}=1.49\text{ GeV}
\end{eqnarray*}
We combine and quote

\[
m_{1}=1.46\pm0.03\text{GeV }
\]
The procedure can be repeated using the general formula
\[
m_{n}^{2}=\frac{\int_{m_{n-1}^{2}}^{R}dt\,t^{2}P_{n+1}(t,R)}{\int_{m_{n-1}^{2}}^{R}dt\,t\,P_{n+1}(t,R)}
\]
with
\begin{equation}
P_{n+1}(t,R).=(1-\frac{t}{m_{n+1}^{2}})(1-\frac{t}{R})
\end{equation}
$R$ being the maximum at which the maximum of $\int_{m_{n-1}^{2}}^{R}dt\,t\,P_{n+1}(t,R)$
takes place (which is also close to the maximum of $\int_{m_{n-1}^{2}}^{R}dt\,t\,^{2}P_{n+1}(t,R)$).

We list our results below

\begin{eqnarray*}
m_{2} & = & m_{N}(1535)=1.47\text{ GeV, \ \ \ \ \ \ \ \ \ \ \ \ \ \ }R=3.2\text{ GeV}^{2}\\
m_{3} & = & m_{N}(1650)=1.57\text{ GeV, \ \ \ \ \ \ \ \ \ \ \ \ \ \ }R=3.5\text{ GeV}^{2}\\
m_{4} & = & m_{N}(1710)=1.74\text{ GeV, \ \ \ \ \ \ \ \ \ \ \ \ \ \ }R=4.0\text{ GeV}^{2}\\
m_{5,6} & = & m_{N}(1880,1895)=1.80\text{ GeV, \ \ \ \ \ \ \ }R=5.3\text{ GeV}^{2}\\
m_{7} & = & m_{N}(2100)=1.98\text{ GeV, \ \ \ \ \ \ \ \ \ \ \ \ \ \ }R=6.4\text{ GeV}^{2}
\end{eqnarray*}
The errors of the predicted masses of resonances is about $\pm10\%$.

\section{Heavy Baryons}

The $\Lambda_{c},\Lambda_{b},\Xi_{c},\Xi_{b}$ and their parity doublets
can be treated by the same sum rule method using an integration kernel
of the form
\begin{equation}
P(t,m^{\prime2})=\left(1-\frac{t}{m^{\prime2}}\right)\left(1-\frac{t}{R}\right)\label{4.1}
\end{equation}
where $m^{\prime}$ is mass of the first exited state or parity doublet
and $R$ is the radius of the circle of Fig.(1).

We first determine the mass of the ground state baryon for a knowledge
of $m^{\prime}$ and then determine $m^{\prime}$ independently and
self-consistently. We consider
\begin{eqnarray*}
 &  & \Lambda_{c}(2286.5)(0,\frac{1}{2}^{+})\ ,\ \Lambda_{c}^{\prime}(2593)(0,\frac{1}{2}^{-})\\
 &  & \Lambda_{b}(5620)(0,\frac{1}{2}^{+})\ ,\ \Lambda_{b}^{\prime}(5912)(0,\frac{1}{2}^{-})
\end{eqnarray*}
In the correlator of Eq.\ref{27} we use the simplest current for
$\Lambda_{Q}$ \cite{Wang}
\begin{equation}
j_{Q}=\frac{1}{\sqrt{2}}\epsilon_{abc}(u_{a}^{T}C\gamma_{5}d_{b}-d_{a}^{T}C\gamma_{5}u_{b})Q_{c})\label{4.3}
\end{equation}
We define as above
\begin{equation}
\Pi(t)=\cancel{q}A(t)+B(t)\label{4.4}
\end{equation}
On the hadronic side we make explicit the contribution of the positive
and negative parity baryons,
\begin{eqnarray}
A(t) & = & -\frac{\lambda_{+}^{2}}{t-m_{+}^{2}}-\frac{\lambda_{-}^{2}}{t-m_{-}^{2}}\label{4.5}\\
B(t) & = & -\frac{\lambda_{+}^{2}m_{+}^{2}}{t-m_{+}^{2}}-\frac{\lambda_{-}^{2}m_{-}^{2}}{t-m_{-}^{2}}\label{4.5a}
\end{eqnarray}
where $\lambda_{\pm}$ the coupling of the current Eq. \ref{4.3}
to the corresponding states.

The QCD spectral functions have been obtained in \cite{Wang} and
are conveniently expressed in \cite{Zhao}.

We define 
\[
\rho_{A}(t)=\frac{3m_{Q}^{4}}{128\pi^{4}}a(t)
\]
with
\begin{eqnarray}
a(t) & = & \int_{m_{Q}^{2}}^{1}dx\ x(1-x)^{2}\left(\frac{t}{m_{Q}^{2}}-\frac{1}{x}\right)^{2}+\frac{\pi^{2}}{6}\frac{\left\langle a_{s}GG\right\rangle }{m_{Q}^{4}}\left(1-\frac{m_{Q}^{4}}{t^{2}}\right)\label{4.6}\\
 &  & -\frac{\pi^{2}}{6}\frac{\left\langle a_{s}GG\right\rangle }{m_{Q}^{4}}\left(1-\frac{m_{Q}^{4}}{t^{2}}\right)^{2}+...\nonumber 
\end{eqnarray}
Consider now the integral $\frac{1}{2\pi i}\int_{C}dt\ A(t)P(t)$
where $C$ is the contour of Fig. 1 in the complex $\ t-$plane.

We claim that our choice of the damping kernel $P(t,R)$ essentially
eliminates all hadronic contributions except that of the ground state
so that
\begin{equation}
\lambda_{+}^{2}P(m_{+}^{2})=\int_{m_{Q}^{2}}^{R}dt\,P(t,R)\rho_{A}(t)=I_{0}\label{4.7}
\end{equation}
Another equation is obtained from the first moment
\[
\lambda_{+}^{2}m_{+}^{2}P(m_{+}^{2})=\int_{m_{Q}^{2}}^{R}dt\,t\,P(t,R)\rho_{A}(t)=I_{1}
\]
$R$ is chosen in the stability region of the integrals $I_{0}$ and
$I_{1}$ for $\lambda_{c}$. These integrals turn out to be very flat
functions of $t$ which obtain a maximum between $4.5^{2}$ and $5m_{c}^{2}$.
The result is

\begin{eqnarray}
m_{\Lambda_{b}} & = & 2.10\text{ GeV}\ \ \ \ \ (\text{exp. }2.23\text{ GeV})\label{4.9a}\\
m_{\Lambda_{b}} & = & 5.92\text{ GeV}\ \ \ \ \ (\text{exp. }5.62\text{ GeV})\label{4.9b}
\end{eqnarray}
Using $B(t)P(t,R)$ yields practically identical results.

We now proceed to calculate $m_{\pm}$ independently. Because $b$
and $c$ quarks are heavy $\Pi(0)$ is given by its QCD expression
which yields an additional sum rule
\begin{equation}
\frac{\lambda_{+}^{2}}{m_{+}^{2}}P(m_{+}^{2})=\int_{m_{Q}^{2}}^{R}\frac{dt}{t}\ P(t,R)\rho_{\Lambda I}(t)=I_{-1}\label{4.10}
\end{equation}

This implies
\[
\frac{I_{1}}{I_{0}}=\frac{I_{0}}{I_{-1}}
\]
from which we can determine $m_{+}$. The result is
\begin{eqnarray}
m_{\Lambda_{c}^{\prime}} & = & (2.51\pm0.03)\text{ GeV}\ \ \ \ \ (\text{exp. }2.58\text{ GeV})\label{4.11a}\\
m_{\Lambda_{b}^{\prime}} & = & (5.87\pm0.14)\text{ GeV}\ \ \ \ \ (\text{exp. }5.91\text{ GeV})\label{4.11b}
\end{eqnarray}
The errors are estimated by varying $R$ by $10\%$.

Finally for the $\Xi$ baryon $\Xi_{Q}(_{5797}^{2468})(\frac{1}{2},\frac{1}{2}^{+}),\ \Xi_{Q}^{\prime}(_{5935}^{2580})(\frac{1}{2},\frac{1}{2}^{+})$
the function $a(t)$ in Eq.\ref{4.6} picks up an additional term
proportional to the strange quark mass $m_{s}$
\begin{equation}
\Delta a(t)=\frac{4\pi^{2}}{3}\frac{m_{s}\left\langle \bar{s}s-2\bar{q}q\right\rangle }{m_{Q}^{4}}\left(1-\frac{m_{Q}^{4}}{t^{2}}\right)\label{4.12}
\end{equation}
The calculation proceeds as before, giving
\begin{eqnarray*}
m_{\Xi_{c}} & = & 2.05\text{ GeV \ \ \ \ \ (exp. }2.468\text{ GeV)}\\
m_{\Xi_{b}} & = & 5.40\text{ GeV \ \ \ \ \ (exp. }5.79\text{ GeV)}\\
m_{\Xi_{c}^{\prime}} & = & 2.46\text{ GeV \ \ \ \ \ (exp. }2.58\text{ GeV)}\\
m_{\Xi_{b}^{\prime}} & = & 5.87\text{ GeV \ \ \ \ \ (exp. }5.935\text{ GeV)}
\end{eqnarray*}

\textbf{Conclusions}: We have calculated the masses of the baryon
recurrences with a new variant of QCD finite energy sum rules. The
only free parameter of the sum rules, the radius of the circle in
the complex t-plane, is fixed by the requirement of stability. The
method works well for all similar systems such as the vector resonances.
The main source of error is the zero width approximation for the resonances.
We have estimated this error by allowing the radius entering the sum
rule to vary by $\pm10\%$. Order $\alpha_{s}$ corrections are included,
order $\alpha_{s}^{2}$ are calculated and found to be negligible.
The sum rule predictions are compared with the experimental numbers
and agreement within the expected accuracy is found. It can be concluded
that QCD is applicable to single resonances and their recurrences.

\textbf{Acknowledgements:} This work was supported in part by the
Alexander von Humboldt Foundation (Germany), under the Research Group
Linkage Programme, and by the University of Cape Town (South Africa).\newpage{}

\end{document}